Spontaneous mechanical and energetic state transitions during *Caenorhabditis elegans* gastrulation

Jiao Miao[1†*], Guoye Guan[1†], Chao Tang[1,2,3*]

1. *Center for Quantitative Biology, Peking University, Beijing, China*
2. *Peking-Tsinghua Center for Life Sciences, Peking University, Beijing, China*
3. *School of Physics, Peking University, Beijing, China*

\* For correspondence: miao.physics@gmail.com (JM); tangc@pku.edu.cn (CT)

† These authors contributed equally to this work.

**ABSTRACT**

Gastrulation, namely cell internalization, is a significant milestone during the development of metazoans from worm to human, which generates multiple embryonic layers with distinct cell fates and spatial organizations. Although many molecular activities are known to facilitate this process, in this paper, we focus on gastrulation of the nematode *Caenorhabditis elegans* and theoretically demonstrate that even a group of cells with only isotropic repulsive and attractive interactions can experience such internalization behavior when dividing within a confined space. As the cell number increases and cell size decreases, the cells contacted to the eggshell become closer to each other along with harder lateral compression, and a cell that internalizes could effectively increase the cell neighbor distance and lower the potential energy of the system. The multicellular structure transits from single- to double-layer spontaneously with bistable states existing from 15- to 44-cell stages, near the gastrulation timing *in vivo*. Specifically, the cells with a larger size or placed near a smaller-curvature boundary are easier to internalize. Actively regulating a few cells' internalizations can make the morphogenesis noise-resistant. Our work successfully recaptures the key characteristics in *C. elegans* gastrulation and provides a rational interpretation of how this phenomenon emerges and is optimally programmed.

**INTRODUCTION**

Metazoan embryogenesis starts from a single fertilized zygote and ends in a vast number of cells with stereotypic spatial patterns and fate specifications (*Dehaan and Ebert, 1964*; *Kumar et al., 2015*; *Farrell et al., 2018*; *Packer et al., 2019*). Those cells constitute multiple types of tissues and organs in an animal, such as intestine, muscle, and skin, which need precise and robust control on both morphogenesis and differentiation. During the rapid cleavage of blastomere, a dynamic process called gastrulation is activated at a specific timing to rearrange the cells into different germ layers, usually including endoderm, mesoderm, and ectoderm (*Stern, 2004*; *Solnica-Krezel and Sepich, 2012*). From the view of geometry, all cells are initially located adjacent to the inner side of the boundary (e.g., eggshell), like a two-dimensional curved surface, then they partially internalize so that the inner and outer layers are formed. Although the underlying regulations and the proportion of cells that move inward vary from species to species, this morphological change is common among all kinds of multicellular animals, including but not limited to nematode, ascidian, fly, frog, fish, chick, mouse, and human (*Houthoofd et al., 2006*; *Swalla, 1993*; *Lemke et al., 2015*; *Heasman, 2006*; *Rohde and Heisenberg, 2007*; *Chapman et al., 2002*; *Tam and Behringer, 1997*; *Ghimire et al., 2021*). As the developmental biologist Lewis Wolpert said, "not birth, marriage or death, but gastrulation which is truly the most important time in your life" (*Vicente, 2015*; *Illman, 2021*).

The eutelic nematode *Caenorhabditis elegans* is a type of tiny transparent worm, which is famous for its highly invariant developmental programs at the cellular level and eventually generates 558 non-identical cells in a hatched larva (*Sulston et al., 1983*). Thus, it's one of the most popular model animals for developmental biology and many regulatory mechanisms of gastrulation have been unraveled using this system (*Corsi et al., 2015*; *Nance et al., 2005*; *Goldstein and Nance, 2020*). During *C. elegans* embryogenesis, the gastrulating movements last for a long period (from 26- to ~300-cell stages) and the cells from different lineages internalize group by group (roughly following the order as E → P4 → MS → D → C → AB) (*Chisholm and Hardin, 2005*; *Nance and Priess, 2002*). *Figure 1A* shows the *C. elegans* embryonic development from 4- to 350-cell stages imaged by 3-dimensional (3D) time-lapse confocal microscopy, which provides quantitative information for each cell's nucleus position (GFP-labeled) and membrane morphology (mCherry-labeled) (*Cao et al., 2020*). Prior to gastrulation onset, all cells are in contact with the shell (*Nance and Priess, 2002*). Around 26- to 28-cell stages, gastrulation starts and the two gut precursor cells, Ea and Ep in E lineage (i.e., E2), together migrate inward coordinated by their intensive adhesion to lateral neighbors and apical contraction near the shell (*Marston et al., 2016*). At 44-cell stage (right before E2 divisions), those motions result in a layered multicellular structure consisting of the E2 cells inside and the others outside (*Figure 1B*). Using the segmented cell morphology of 17 *C. elegans* wild-type embryos (*Cao et al., 2020*), the total number of internal cells (without contact with the outer space) is estimated, showing that the first fully internalization occurs at 42- to 46-cell stages and the internal cells occupy around 18 ± 1% of all cells in a ~350-cell embryo (*Figure 1C-D* and *Table S1*).

Despite that gastrulation is well studied in morphological and molecular levels, a comprehensive mechanical and energetic interpretation of such a general phenomenon (hereafter referred to as cell internalization) is lacking. The evolutional origin, emergence principle, and program strategy of gastrulation are largely unclear (*Nakanishi et al., 2014*). Previous reports showed that for different nematode species, cell adhesion essentially affects the 4-cell arrangement, while the curvature of the eggshell seems to be associated with the gastrulation timing (*Yamamoto and Kimura, 2017*; *Schierenberg, 2006*). How the physical parameters play their roles in cell internalization, including its activation timing, and the number and identity of internalizing cells, remains to be answered.

In this paper, we establish a computational pipeline to simulate *C. elegans* embryogenesis from 4- to 330-cell stages, using a coarse-grained model that simplifies the proliferating cells as particles interacting with isotropic repulsion and attraction. We find that cell internalization occurs spontaneously in all conditions of physical parameter setting, during which the cells are rearranged more sparsely and the system's potential energy becomes lower. Over the blastomere cleavage, the compressive pressure between cells is enhanced and a more stable state with cell internalized emerges when the total cell number exceeds a threshold. Surprisingly, with the physical parameters obtained from the real embryo, the predicted timing of cell internalization is close to the one observed experimentally. Moreover, we systematically explore how different factors affect cell internalization, including stiffness and adhesion of cells, cell size, cell position, etc. Theoretically, the larger cell is easier to internalize, and the first internalizing cells E2 in a living embryo are also the largest due to their genetically regulated cell cycle lengthening. The match between model prediction and experimental observation suggests that the regulatory programs of gastrulation are designed considering the fundamental mechanical and energetics laws. Last but not least, we investigate how cell internalization responds to motional noise and the stabilizing effect of the active force that drives specific cells' internalization.

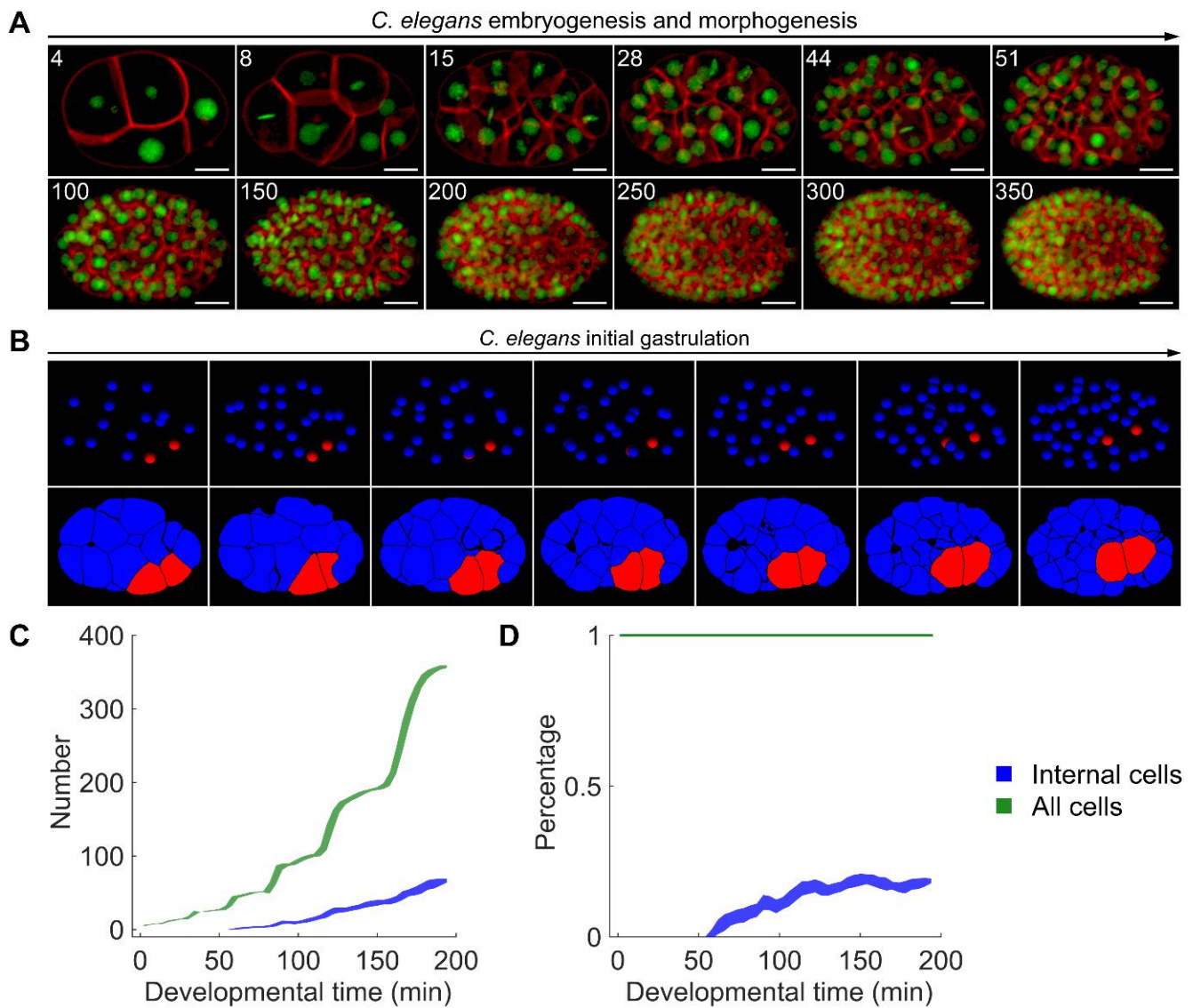

**Figure 1.** The cell internalization during *C. elegans* early development (data obtained from *Cao et al., 2020*). (A) *C. elegans* embryogenesis and morphogenesis imaged by 3D time-lapse confocal microscopy; green fluorescence, cell nuclei labeled by GFP; red fluorescence, cell membranes labeled by mCherry; from top left to bottom right, image projection along the shooting direction when the total cell number = 4, 8, 15, 28, 44, 51, ~100, ~150, ~200, ~250, ~300, and ~350 as denoted on the top left of each subgraph; scale bar, 10 μm. (B) Nuclei tracking and membrane segmentation (cross-section image of a specific focal plane) at selected time points for illustrating the initial gastrulation; the 1st row, nuclei tracking results; the 2nd row, membrane segmentation results; the interval between consecutive images is 4 time points ≈ 5.6 min; the gut precursor cells E2 and the other cells are colored in red and blue respectively. (C, D) An estimation on (C) the number and (D) the percentage of all cells and internal cells over time using 17 wild-type embryos recorded from 4- to 350-cell stages; the painted region represents the data distribution by mean value ± standard deviation; the starting point of developmental time is set by the last moment of 4-cell stage; blue, internal cells without contact with the outer space; green, all cells with a detected nucleus.

**RESULTS**

***A computational pipeline based on a coarse-grained model is established for simulating cell-cell interactions, cell divisions, and cell motions within a confined space.***

A coarse-grained model, in which cells are simplified as particles and subject to distance-dependent repulsion and attraction from their neighbors and repulsion from the shell, was proposed to be capable to produce the multicellular structures that resemble the ones observed in the live *C. elegans* embryo (*Figure 2A-B*) (*Fickentscher et al., 2013*; *Fickentscher et al., 2016*; *Yamamoto and Kimura, 2017*). In this model, the cell motion is constrained inside an ellipsoidal shell which is axisymmetric about its anterior-posterior axis, namely the major axis along *x* (*Figure 2C*). The lengths of the semi-major and semi-minor axes are $L_x$ = 27 μm and $L_y = L_z$ = 18 μm respectively. All the cell division orientations are set nearly along the anterior-posterior axis and tangential to the shell, with the exception that the ABa and ABp cells divide along the left-right axis, namely the minor axis along *y* (*Figure 2D*; see *Materials and Methods*) (*Sugioka and Bowerman, 2018*). Such programs can minimize the effect of cell division orientation on cell internalization. The mechanical parameters (i.e., eggshell stiffness, cell stiffness, cell adhesion) have been well-tuned according to the experimental structures at 4-, 8-, 12-, 26-, and 51-cell stages in our previous work (*Guan et al., 2020*). Specifically, the stiffness (repulsive factor *K*) and adhesion (attractive factor *α*) are assumed as homogeneous for all the cells and stages during embryogenesis. The geometric parameters such as cell volume segregation ratio and shell size are assigned with the values measured by *in vivo* experiments and detailed in *Materials and Methods*. Unless otherwise specified, the system is set as noise-free. The simulation starts from the rhombic 4-cell structure and proceeds according to the highly conserved cell division sequence recorded experimentally until reaching 330 cells (*Table S2*) (*Guan et al., 2019*). Such a reductionist mechanical model is expected to recapitulate the most fundamental features and rules that govern *C. elegans* embryonic morphogenesis.

***Cell internalization takes place spontaneously in all conditions of physical parameter setting.***

In the simulation with experimental parameters (*K* = 0.01 and *α* = 0.9; Condition 1), all cells contact with the shell until 46-cell stage. When the total cell number reaches 46, a cell in E lineage moves toward the embryo center from the periphery, followed by more and more cells internalizing at later stages (*Figure 2E-F*, *Table S3*, and *Videos S1-S2*). Intriguingly, the developmental stage when the initial cell internalization occurs is close to the one in the real embryo. During *C. elegans* gastrulation, the E2 cells start ingressing around 26- to 28-cell stages, settle and divide at 44-cell stage, afterward the blastomere has 46 cells (*Figure 1B*) (*Nance and Priess, 2002*; *Nance et al., 2005*). To test how much the cell internalization depends on the experimental parameters, we consider 5 more conditions with different physical parameters settings, by changing the attractive force, division asymmetry, division orientation, and division sequence from the experiment-based default to the simplest case (i.e., without attractive force and with symmetric division whose orientation and sequence are fully randomized). Regarding the conditions with stochasticity, a total of 100 independent repeats are generated for each condition and their mean behavior would be used for subsequent comparison. Surprisingly, cell internalization emerges in all the conditions even though the activation timing varies from 44- to 87-cell stages, indicating the spontaneity of such phenomenon and its high independence on the physical parameters considered here (*Figure 2G-H* and *Table S3*).

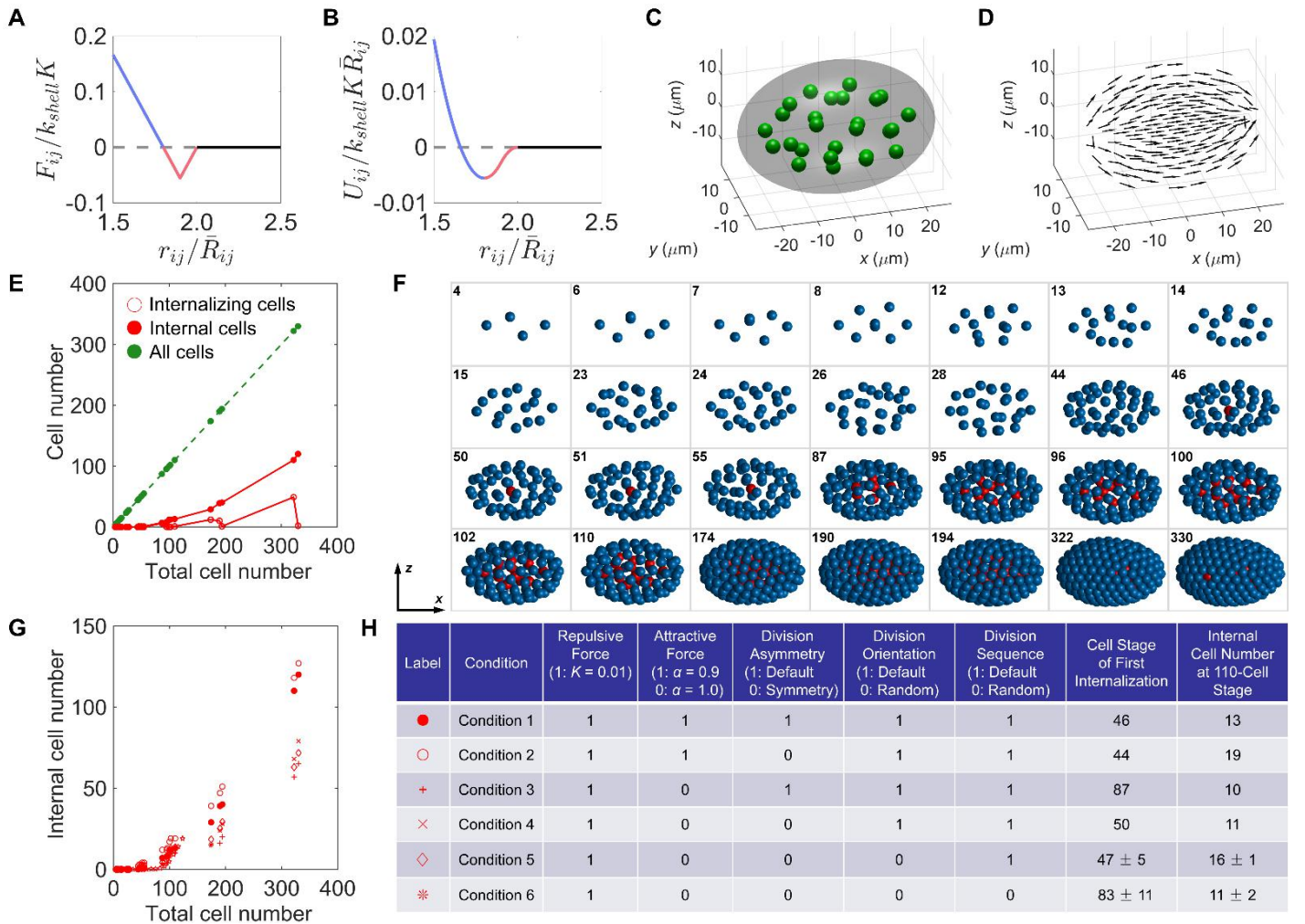

**Figure 2.** The simulation on *C. elegans* early embryogenesis and morphogenesis. (A, B) Schematics of dimensionless (A) force $\frac{F_{ij}}{k_{shell}K}$ and (B) potential $\frac{U_{ij}}{k_{shell}K\bar{R}_{ij}}$ between cells, which rely on the dimensionless cell-to-cell distance $\frac{r_{ij}}{\bar{R}_{ij}}$ (*i* and *j* represent two cells' identity labels; $F_{ij}$, $U_{ij}$, and $r_{ij}$ represent the original force, potential, and distance between two cells, respectively; $k_{shell} \equiv 1$ and $K$ represent the stiffness of the shell and the cell respectively; $\bar{R}_{ij}$ represents the average of two cells' radii); blue line, repulsion when $\frac{r_{ij}}{\bar{R}_{ij}} < 2\alpha$; red line, attraction when $2\alpha \leq \frac{r_{ij}}{\bar{R}_{ij}} < 2$; black line, no interaction when $\frac{r_{ij}}{\bar{R}_{ij}} \geq 2$. (C) Schematic of cells placed inside an ellipsoidal shell in mechanical equilibrium; green sphere, cells; gray shadow, shell. (D) Schematic of cell division orientations set nearly along the anterior-posterior axis and tangential to the shell; the arrow denotes the prescribed cell division orientation at a specific position. (E) Number of internalizing cells and internal cells over the increase of total cell number; red empty circle, internalizing cells; red solid circle, internal cells; green solid circle, all cells. (F) Simulated embryo structures from 4- to 330-cell stages (from top left to bottom right); the total cell number is denoted on the top left of each subgraph; the cells with and without contact with the shell are colored in blue and red respectively. (G, H) Number of internal cells over the increase of total cell number when the physical parameter settings are different; the 6 conditions and their corresponding symbols and settings are listed in (H). Regarding the conditions with stochasticity (Condition 5 and Condition 6), the cell number in (G) is presented by mean value and the cell stage and cell number in (H) are presented by mean value ± standard deviation.

***Cell internalization can reduce a blastomere's system potential energy by redistributing the cell positions more sparsely.***

Since the emergence of cell internalization relies little on the physical parameter settings (*Figure 2G-H*), next we seek to unveil the basic mechanism of this phenomenon. Here we introduce a phenomenological model, which is further simplified into 2-dimensional (2D) and consists of an oval-shaped shell ($L_x$ = 27 μm, $L_y$ = 18 μm, $\bar{L} = \sqrt{L_x L_y}$) and $N$ cells inside. Those cells have an area equally divided from the shell's area and interact with each other only through repulsion ($K$ = 0.01 and $\alpha$ = 1). They are placed internally tangent to the boundary and have the same distance to their neighbors. Then we construct the single-layer structure (with $N$ cells contacting the boundary) and double-layer structure (with $N$-1 cells contacting the boundary and 1 cell at the center) to represent the topological change before and after cell internalization (*Figure 3A*). Interestingly, when the total cell number is small, the single-layer structure has smaller system potential energy and longer average neighbor distance than the ones of double-layer structure (*Figure 3B-C*). Thus, all the cells are located adjacent to the shell, obeying the minimum total potential energy principle. The relationship changes as the total cell number grows, explaining why the *in silico* blastomere tends to be single-layer in the beginning and double-layer in the end. This result suggests that the single-layer structure becomes more and more compressed in the lateral direction and the stress and energy can be released by a structural change. In the simulation with experimental parameters, the first cell internalization proceeds between time point $10^4 \sim 10^5$, also accompanied by a decrease in system potential energy and an increase in average neighbor distance, supporting the explanation that the cells are repositioned sparsely and the system releases its potential energy by cell internalization (*Figure 3D-E* and *Table S4*).

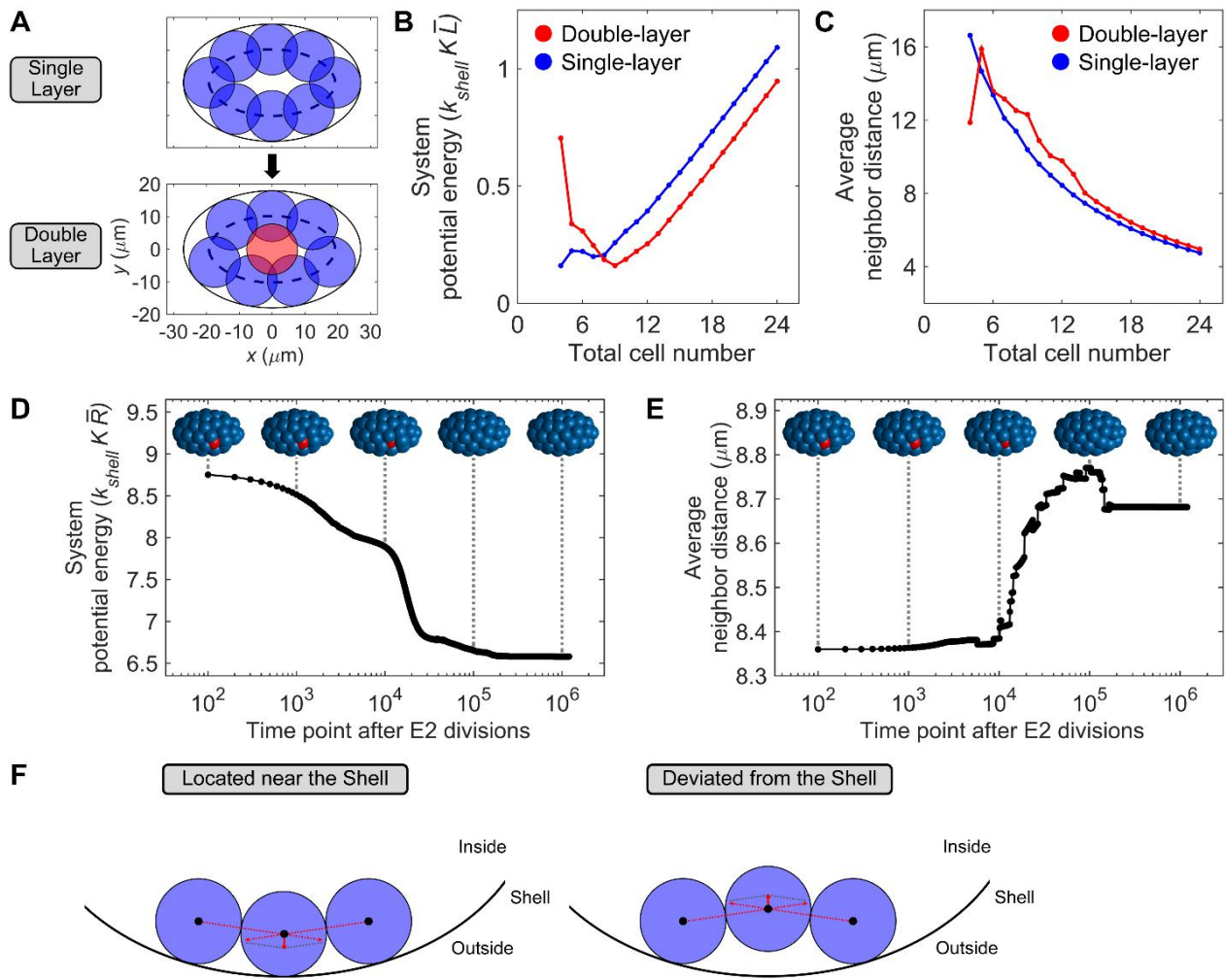

**Figure 3.** The mechanical and energetic explanation of cell internalization. (A) A 2-dimensional system combining an elliptical shell (long diameter $2L_x = 2 \times 27$ μm, short diameter $2L_y = 2 \times 18$ μm) and $N$ cells with an equal area $\frac{\pi L_x L_y}{N}$; the single- and double-layer structures are illustrated on top the bottom respectively; the cells with and without contact with the shell are colored in blue and red respectively. (B) The change of system potential energy as the total cell number grows in the 2D system; $\bar{L} = \sqrt{L_x L_y}$ is the equivalent radius of the circle with the same area as the elliptical shell. (C) The change of average neighbor distance as the total cell number grows in the 2D system. (D, E) The temporal change of (D) system potential energy and (E) average neighbor distance in the simulation with experimental parameters when the first cell internalization proceeds (46-cell stage); the embryo structures at time points $10^2$, $10^3$, $10^4$, $10^5$, and $10^6$ are illustrated on top, with the internalizing cell painted in red and the others painted in blue. The system potential energy in (B, D) is nondimensionalized by using $k_{shell}K\bar{L}$ and $k_{shell}K\bar{R}$ as a unit respectively, where $\bar{R} = \sqrt[3]{\frac{\sum_{i=1}^{N} V_i}{N}}$ and $V_i$ denotes the cell $i$'s volume. The neighborhood in (E) is defined when the contact area is larger than 5% of a cell's surface area for stability and reliability, and only the space within the ellipsoidal shell is considered (*Rycroft, 2009*). (F) The resultant force imposed on a specific cell by its two neighbors when it is located near the shell (left) or deviated from the shell (right); the component and resultant repulsive forces are illustrated by solid and dashed arrows respectively.

### *Cell internalization is primarily driven by cell-cell repulsion but not attraction.*

Using the phenomenological model, we perform force analysis onto the cell which contacts with two neighbor cells and a curved boundary. At an initial state, the resultant force imposed on a cell by its neighbors is oriented towards the shell due to the curvature. A restoring force could exist when the cell's position is slightly perturbed, making it stably contact the shell with pressure (*Figure 3F*). When the perturbation is strong enough and makes the cell more distant from the shell, the resultant force from neighbors would turn to be oriented inward and no restoring force remains. Therefore, a cell would keep ingressing irreversibly once it leaves the position of equilibrium. To validate the hypothesis that cell-cell repulsion is key for a cell to internalize, we adopt the simulation with experimental parameters and eliminate the attractive force between the first internalizing cell and its neighbors. This modification results in internalization of the same cell and similar temporal patterns in both system potential energy and average neighbor distance, supporting the repulsive force as the key of cell internalization instead of attractive force (*Figure 3D-E*, *Figure S1* and *Table S4*).

Regarding the force field with both intercellular repulsion and attraction, the attraction, which mainly origins from adhesive proteins and gap junctions, neutralizes a part of the repulsion (associated with cytoskeleton) all across the effective range $0 < \frac{r_{ij}}{\bar{R}_{ij}} < 2$ and changes the cell-to-cell distance in mechanical equilibrium from $r_{ij} = 2\bar{R}_{ij}$ to $r_{ij} = \alpha \cdot 2\bar{R}_{ij}$ ($\alpha < 1$) (*Figure 4A-B*) (*Marston et al., 2016*; *Simonsen et al., 2014*; *Pegoraro et al., 2017*). Therefore, the attraction weakens the outward-oriented repulsive force on a cell and increases the chance of cell internalization, while repulsion has the opposite effect. To verify this idea, we add motional noise into the simulation with experimental parameters and change the repulsive and attractive strengths globally ($\eta = 10^{-4} \sim 10^{-2}$, $K_R = 10^{-3} \sim 10^{-1}$, $K_A = 10^{-3} \sim 10^{-1}$, repeat number = 50) (see *Materials and Methods*). Here, $\eta$ is the amplitude of motional noise; $K_R$ and $K_A$ are defined as the slope of the $\frac{F_{ij}}{k_{shell}K} - \frac{r_{ij}}{\bar{R}_{ij}}$ curve in the repulsion range $0 < \frac{r_{ij}}{\bar{R}_{ij}} < 2\alpha$ and attraction range $2\alpha \leq \frac{r_{ij}}{\bar{R}_{ij}} < 2$ respectively, and the curve has a fixed point at $\frac{r_{ij}}{\bar{R}_{ij}} = 2\alpha$ (*Figure 4C*). As expected, the activation timing (i.e., total cell number) of the first cell internalization is positively correlated to the repulsive strength and negatively correlated to the attractive strength, suggesting that the repulsive and attractive force can inhibit and enhance the trend of cell internalization respectively (*Figure 4D-E* and *Table S5*).

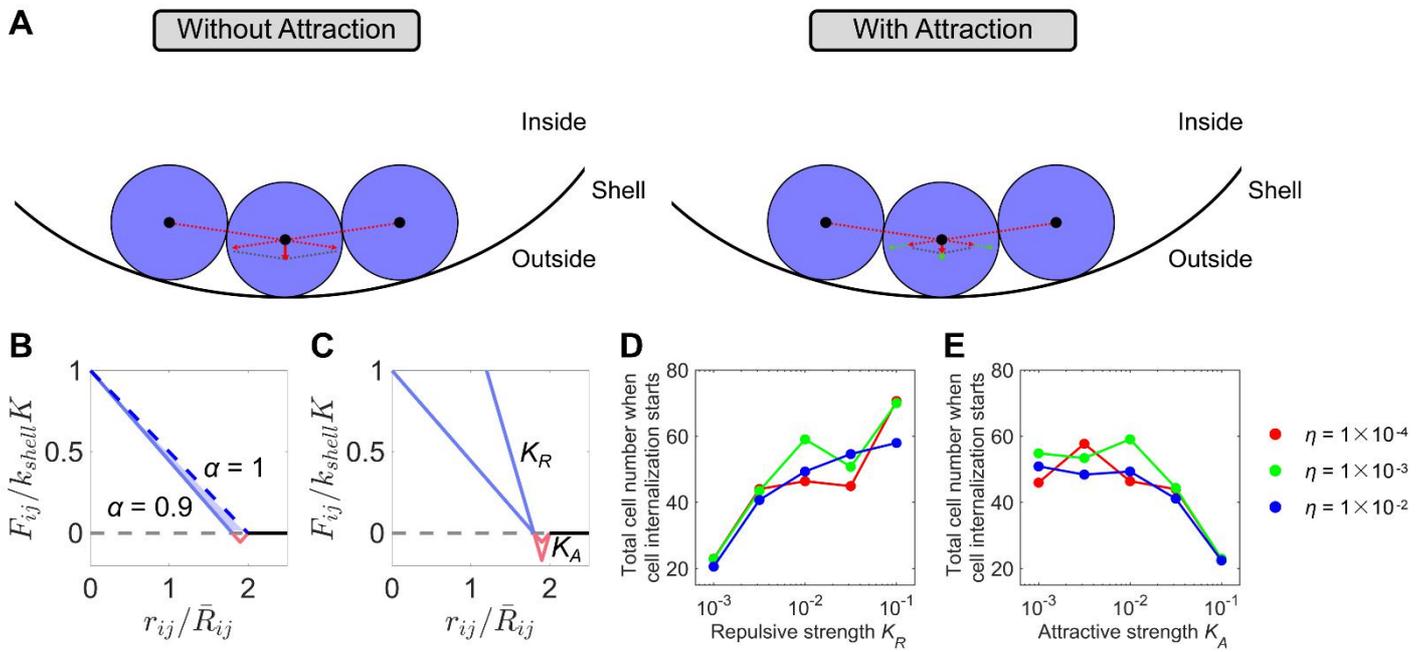

**Figure 4.** The effect of cell-cell attraction on cell internalization. (A) The resultant force imposed on a specific cell by its two neighbors when there's only repulsive force (left) or both repulsive and attractive force (right); the component and resultant forces are illustrated by solid and dashed arrows respectively; red, repulsive force; green, attractive force. (B) Schematics of dimensionless force $\frac{F_{ij}}{k_{shell}K}$ between cells; solid line, force field consisting of both repulsion and attraction ($\alpha = 0.9$); dashed line, force field consisting of only repulsion ($\alpha = 1$); the repulsion neutralized by attraction is highlighted by blue shade. (C) Schematics of dimensionless force $\frac{F_{ij}}{k_{shell}K}$ between cells when the repulsive strength $K_R$ and attractive strength $K_A$ vary from $10^{-3}$ to $10^{-1}$. The meanings of coefficient and color in (B, C) are the same as the ones in *Figure 2A-B*. (D, E) The total cell number when cell internalization starts under different value assignments on global (D) repulsive strength and (E) attractive strength; the relationship between colors and noise amplitudes is denoted on right.

***The blastomere becomes bistable in single- and double-layer structures during a developmental stage close to the in vivo gastrulation timing.***

To detect the possible equilibrium states of the blastomere, we artificially push Ea or Ep cell (the cells firstly internalizing *in vivo*) inward at 14-, 15-, 23-, 24-, 26-, 28, and 44-cell stages respectively and record the changes of system potential energy during their driven internalization. The active driving force would be eliminated when the cell is away from the shell to check if its internalization can be stabilized (see *Materials and Methods*). Before 15-cell stage, the system with a cell inside is unstable as the cell pushed inward moves back to the outside once the force is removed (*Figure 5A*). As the cells continue to divide, a new stable state in which at least one cell can maintain internalized emerges. The energy of the double-layer structure ($E_d$; after cell internalization) is smaller than the single-layer one ($E_s$; before cell internalization), with an energy barrier between them ($E_b$) and hampering the structural transition (*Figure 5B* and *Table S5*). Both the decreases of $E_d - E_s$ and $E_b - E_s$ over the total cell number suggest that the double-layer structure is stabler and stabler than the single-layer one and the transition is easier and easier. As $E_b - E_s$ approaches 0 at 46-cell stage, the energy barrier vanishes so that the structural transition becomes completely spontaneous and no active force is needed (*Figure 3D*). Such transitions from single- to double-layer structure are attributed to the consecutive cell divisions which lead to the shorter distance and stronger compression between the outer cells.

The first gastrulation in a real *C. elegans* embryo starts at 26-cell stage and ends at 44-cell stage, near the emergence of bistability predicted by the simplified model (i.e., 15- to 44-cell stages), suggesting that the genetic programs are designed following the mechanical and energetic laws and utilizing the intercellular compressive pressure that increases over cleavage and lowers the energy barrier. Nevertheless, the active regulation (e.g., adhesion and contractility) plays a role in helping the system get over the energy barrier and directing the structural transition (*Farhadifar et al., 2007*).

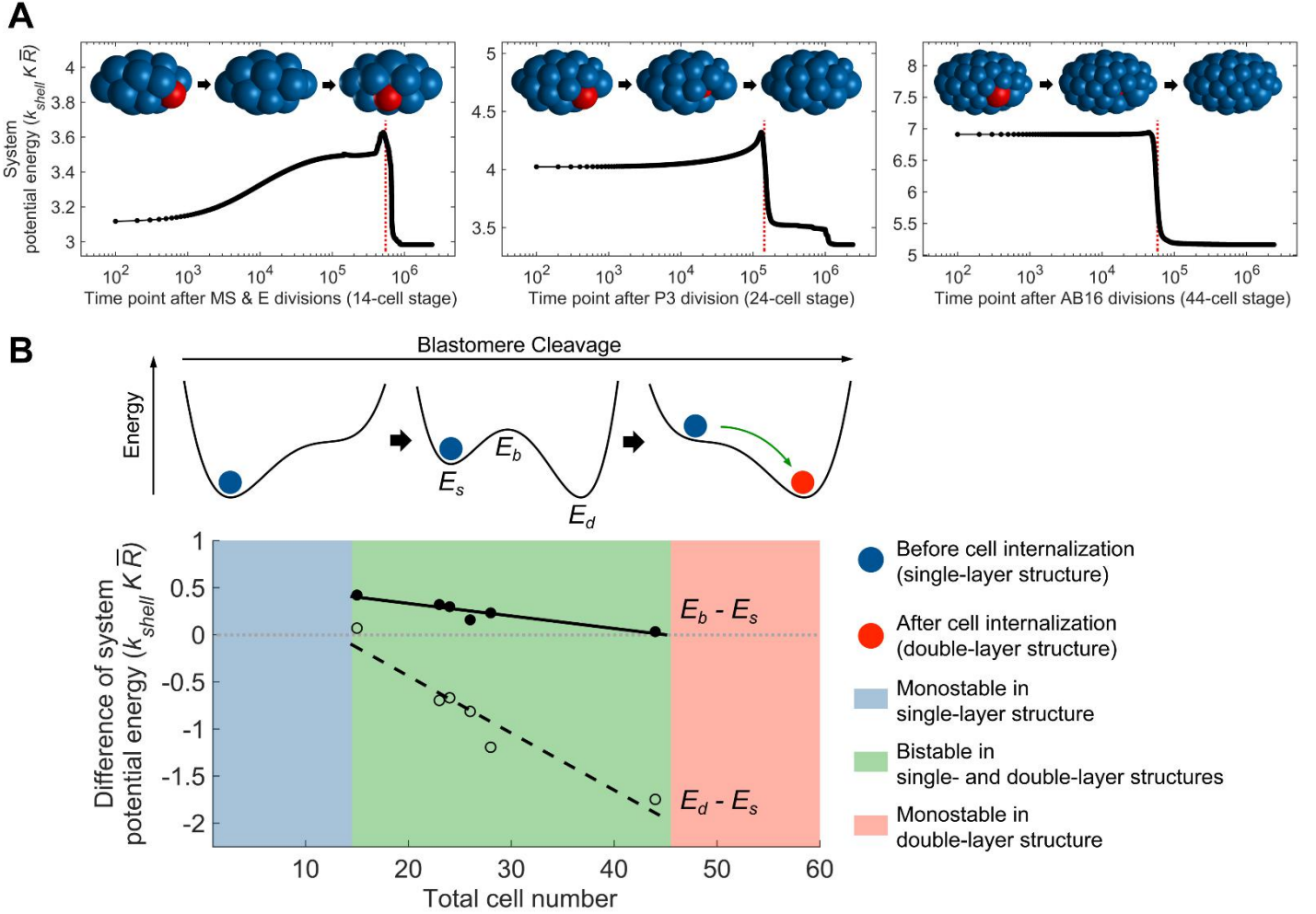

**Figure 5.** The monostable and bistable states that transit during blastomere cleavage. (A) The temporal change of system potential energy in the simulation with experimental parameters and an active driving force imposed on Ep cell at 14-, 24-, and 44-cell stages; the embryo structures at time point $10^2$, the moment after Ep internalization and elimination of its active driving force (dashed red line), and time point $2.4 \times 10^6$ are illustrated on top from left to right, with the internalizing cell painted in red and the others painted in blue. (B) Difference of system potential energy between single-and double-layer structures over blastomere cleavage; the schematics of three conditions corresponding to the simulation results under different total cell numbers (shaded with blue, green, and red) are illustrated on top; $E_s$ and $E_d$ represent the system potential energy of single- and double-layer structures respectively while $E_b$ is the energy barrier between them. The system potential energy in (A, B) is nondimensionalized by using $k_{shell} K \bar{R}$ as a unit respectively, where $\bar{R} = \sqrt[3]{\frac{\sum_{i=1}^{N} V_i}{N}}$ and $V_i$ denotes the cell $i$'s volume.

***The cells with a larger size or placed near a smaller-curvature boundary have a higher probability to internalize.***

Here, we explore which kinds of cells are easier to internalize. For simplicity, we utilize the 2D phenomenological model and the 3D mechanical simulation with only intercellular repulsion, symmetric division, and randomized division orientation and sequence considered (Condition 6, *Figure 2H*). The simplest scene can clear up the effect of minor factors and reveal the most fundamental rules on cell internalization.

First, when a cell has a substantially larger size than its two neighbors, the direction of the resultant force imposed on it would change from outward to inward (*Figure 6A*). In the simulation, as the cells are programmed to divide equally round by round, there are only two kinds of cells with different sizes (the larger has a volume twice of the smaller). Hence, we classify the cells into "larger cells" and "smaller cells" from 1- to 128-cell stages; if the total cell number is the power of 2, the data is excluded because all the cells are homogeneous. The simulation (repeat number = 100) shows that the larger cells indeed have a higher probability of cell internalization than the smaller cells (*Figure 6B*). Fascinatingly, in the real *C. elegans* embryo, the first internalizing cells, E2, acquire the largest volume among all cells by essentially postponing their divisions, which is induced by tissue-specific zygotic transcription and gap phase introduction (*Figure 6C-D*) (*Wong et al., 2016*). Given that the lengthened cell cycle in E2 is regulated by a bunch of genes and its defect is associated with gastrulation failure, those mechanisms may be programmed to utilize the simple but significant mechanical effect of inhomogeneous cell size on facilitating cell internalization (*Knight and Wood, 1998*; *Lee et al., 2006*; *Sullivan-Brown et al., 2016*).

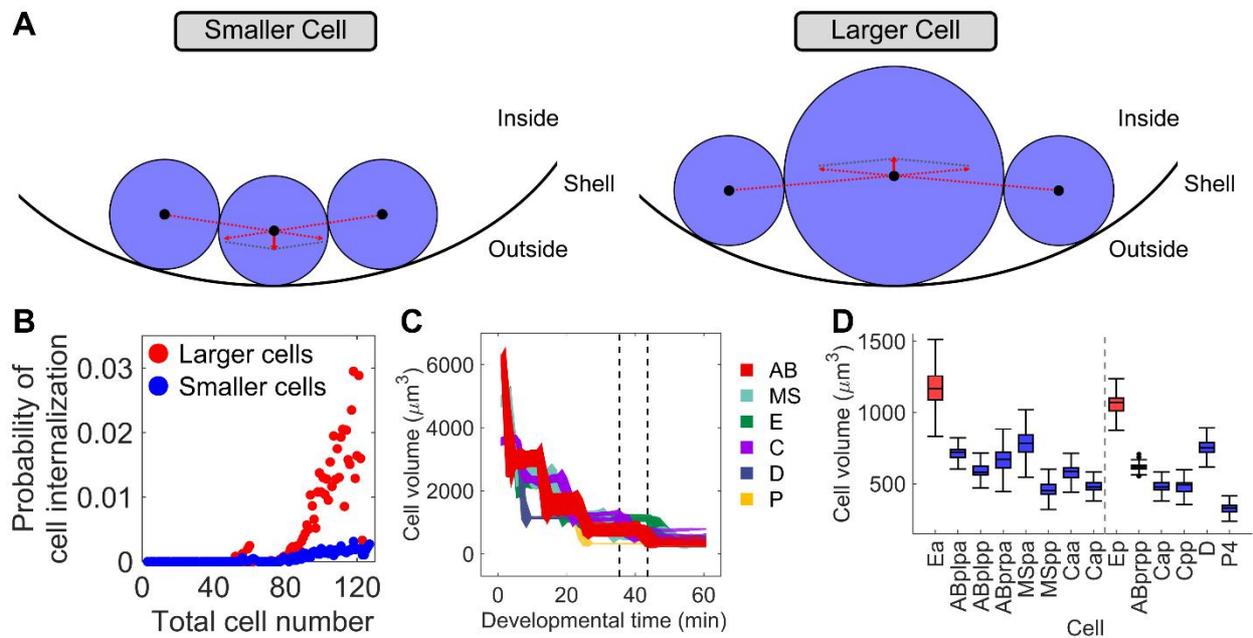

**Figure 6.** The effect of cell size on cell internalization. (A) The resultant force imposed a specific cell by its two neighbors when it has a relatively smaller (left) or larger size (right); the component and resultant repulsive forces are illustrated by solid and dashed arrows respectively. (B) Probability of cell internalization in the simulation with only intercellular repulsion, symmetric division, and randomized division orientation and sequence; red points, larger cells; blue points, smaller cells. (C) Change of cell volume in different cell types over developmental time; the starting point of developmental time is set by the last moment of 4-cell stage; the start and end of the first cell internalization, namely 26- to 44-cell stages, are denoted by black dashed lines; the relationship between colors and cell types (lineage origins) is denoted on right. (D) Cell volume of E2 cells and their neighbors at 28-cell stage; left panel, Ea cell and its neighbors; right panel, Ep cell and its neighbors; for the data of each cell, a colored box is used to illustrate the range between the lower and upper quartiles, along with a black line inside indicating the median and two bars representing the lower and upper inner fences defined by lower quartile - 1.5× interquartile range and upper quartile + 1.5× interquartile range; the mild outliers beyond the bars are shown by black points. The measurements in (C, D) are performed using the data of 17 segmented embryos collected in *Cao et al., 2020*.

Secondly, when a cell is placed near the boundary with a smaller curvature, the forces imposed on it by its similar-sized neighbors would have less weight outward, as reported before (*Figure 7A*) (*Shinbrot et al., 2009*). For the ellipsoidal shell in our study, the curvature is small in the middle and large on both sides, suggesting that the cells around the center are easier to internalize. In the simulation, distributions of both the larger and smaller internalizing cells obey the bell shape, supporting this hypothesis (*Figure 7B-C*). In the real *C. elegans* embryo, a total of 66 cells were proposed to gastrulate (*Harrell and Goldstein, 2011*), among which 32 cells have full lifespan continuously recorded in *Cao et al., 2020*. Here, we take an average of a cell's positions at its first and last time points to approximate its internalizing position (*Table S1*). Regarding the 32 cells, the 1/3 outermost region ($x > 18$ μm or $x < -18$ μm) contains less than 8% data, while the 1/2 central region ($-13.5$ μm $< x <$ 13.5 μm) contains over 74% data (*Figure 7D*). The bell shape calculated with real cells is wider than the ones in simulation, implying that other unknown cues may exist and affect the cell internalization programs as well. Given that the embryo *in vivo* is artificially compressed for better imaging quality, the difference between *in silico* and *in vivo* shell shapes may be a reason and merits further study.

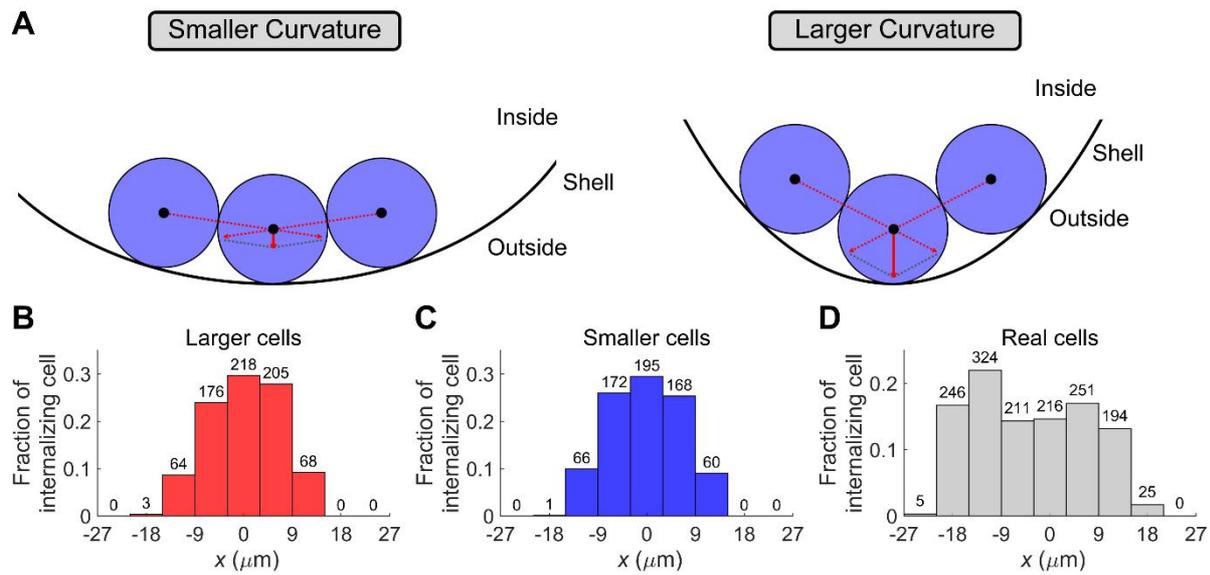

**Figure 7.** The effect of shell curvature on cell internalization. (A) The resultant force imposed on a specific cell by its two neighbors when placed near a smaller-curvature (left) or larger-curvature boundary (right); the component and resultant repulsive forces are illustrated by solid and dashed arrows respectively. (B, C) Fraction of internalizing cells of the (B) larger and (C) smaller cells in the simulation with only intercellular repulsion, symmetric division, and randomized division orientation and sequence; the internalizing cell number in each region is labeled on the top of the column. (D) Fraction of internalizing cell in the real embryo; the internalizing cell number in each region is labeled on the top of the column.

***Cell internalization behaviors could be disturbed by motional noise and stabilized by active driving force imposed on specific cells to regulate their inward movements.***

During metazoan gastrulation, the activation timings and cell types that move inward are crucial because those cells with specific fates must play their roles after internalization, for instance, generating the endoderm or mesoderm of the gastrula. However, the motional noise always exists during embryogenesis (*Li et al., 2018*; *Guan et al., 2019*). It brings variability in the cell positions among individuals and is harmful to the stereotypic developmental programs, such as the location and communication of specific cells (*Sulston et al., 1983*; *Eisenmann, 2005*; *Priess, 2005*). Besides, the accurate timings of cell division and migration are critical to both the precision and robustness of cell arrangement in the embryo (*Fickentscher et al., 2016*; *Tian et al., 2020*). For the E2 cells in *C. elegans* embryo, which are the only precursors of the gut and the first internalizing cells, active regulations (e.g., asymmetric adhesion and apical contraction) are programmed to drive their gastrulating movements at a fixed time (*Marston et al., 2016*).

To elucidate the effect of global motional noise and local active force, we exert a white noise onto all cells' movements and a force pointing inward onto both the E2 cells to mimic their migratory regulation *in vivo* (see *Materials and Methods*) (*Marston et al., 2016*). For each noise amplitude $\eta$ ($2.5 \times 10^{-6} \sim 1.0 \times 10^{-2}$) and force intensity $\kappa$ ($0 \sim 1$), the system is simulated 50 times to obtain its variability in different developmental properties when the first cell internalization is completed (*Table S7*). Notably, the internalizing cells always belong to E lineage when $\eta \leq 1.0 \times 10^{-5}$, and start to be derived from other lineages when $\eta > 1.0 \times 10^{-5}$; the cell type conservation decreases about 50% when $\eta$ reaches $2.5 \times 10^{-4}$ (*Figure 8A*). On the other hand, a driving force with an intensity approaching the level of intercellular repulsion (i.e., $\kappa = 1.0$) can rescue the cell type accuracy of internalizing cells to over 80% even when $\eta$ reaches $1.0 \times 10^{-2}$ (*Figure 8B*). In addition, compared to the non-regulated system, the regulated ones can substantially reduce the variability of the total cell number and internalizing cell number when the first internalization takes place (*Figure 8C-D*). In other words, the precision and robustness of this morphogenetic process can be substantially enhanced by selecting a limited set of cells, instead of all cells, for migratory control, which could serve as a strategy to save system resources (*Table S7*).

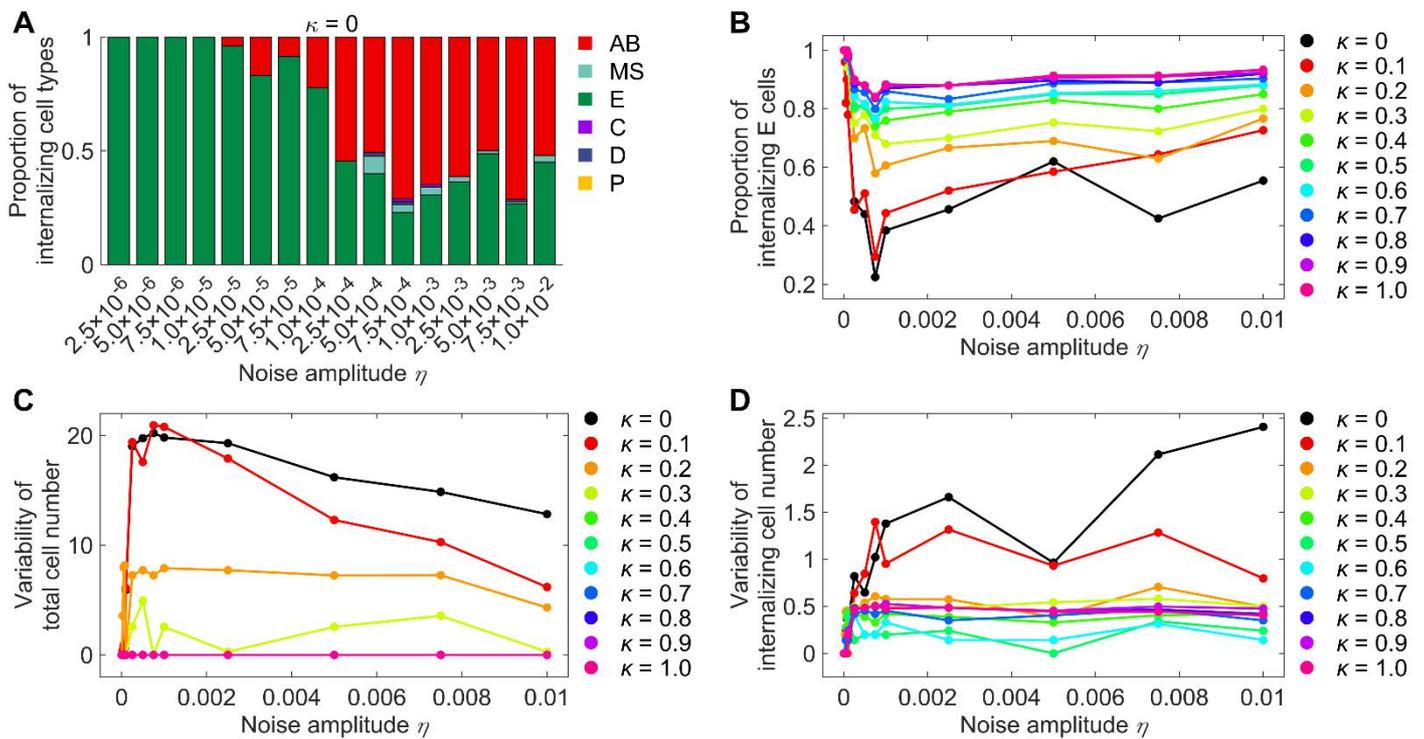

**Figure 8.** The simulation on the first cell internalization under different amplitudes of motional noise and intensities of active force. (A) Proportion of internalizing cell types when $\kappa = 0$; the relationship between colors and cell types is denoted on right. (B) Proportion of internalizing E cells. (C) Variability of total cell number. (D) Variability of internalizing cell number. The relationship between colors and $\kappa$ values is denoted on the right of (B, C, D).

## DISCUSSION

Gastrulation, namely cell internalization, is a significant morphogenetic event for a developing embryo to generate distinct germ layers and is common among the animal kingdom. This work focuses on the stereotypic developmental programs of *C. elegans* and utilizes a coarse-grained model to simulate the cell-cell interaction, cell division, and cell motion during its embryogenesis. We demonstrate that cell internalization can still emerge even though the system is simplified as a spatially-confined proliferating cell aggregate with only homogeneous intercellular repulsion and attraction. Over the cleavage, cell size and cell-cell distance decrease, thus the cell suffers an increasing pressure from its neighbors and a slight perturbation could make the resultant force inward the embryo, leading to spontaneous cell internalization. After the topological change, the cell neighbor distance is increased and the system potential energy is decreased. Furthermore, energy analysis reveals that with the cell number increases, the embryonic structure is initially monostable in single-layer, then becomes bistable in both single- and double-layer, and ends in monostable in double-layer. During the bistable stages (15- to 44-cell stages), the embryo is possible to turn into either structure if it suffers motional noise and there's no extra regulation, explaining why the gastrulation is genetically programmed in a close time window *in vivo* (26- to 44-cell stages) (*Marston et al., 2016*). The active control on specific cells' internalization is fail-safe for morphogenesis in all the aspects of space, time, and cell fate pattern.

A 2D phenomenological model, together with kinematics simulation, reveals that cell stiffness and cell adhesion enhance and inhibit the trend of cell internalization respectively. Besides, the cells with a larger size or placed near a smaller-curvature boundary are easier to internalize. Nonetheless, other biophysical factors (e.g., embryo volume, shell shape, division orientation) plausibly influence cell internalization as well. Exemplified with the embryo volume, it's found that the total volume of cells keeps declining since gastrulation onset in the real embryo (*Figure S2A-B*). We adopt the simulation with experimental parameters, add the motional noise, and assign different embryo volumes at the beginning. As the volume occupancy rate increases, the final number of internal cells at 330-cell stage decreases ($\eta = 10^{-4} \sim 10^{-2}$, embryo volume/shell volume = 0.8 ~ 1.0, repeat number = 50) (*Figure S2C* and *Table S8*). Why the embryo volume decreases over development and if it's involved with gastrulation are unclear. The potential function of such factors is worth an in-depth investigation.

The simplified coarse-grained model well-tuned for *C. elegans* embryogenesis is extensible and capable for study on other forms of tissue- or cell-specific regulations, in addition to the E2-specific active driving force. In the simulation with experimental parameters and motional noise, a few internal cells are found to move outward stochastically, severely harming the embryo's structural accuracy ($\eta = 2.5 \times 10^{-6} \sim 1.0 \times 10^{-2}$, repeat number = 50) (*Figure S3* and *Table S9*). Apparently, one can trap those cells inside the *in silico* blastomere by strengthening the attraction between the internal cells, and such regulation may have been employed *in vivo* as it was previously reported that adherence junctions would form in the intestine (*Leung et al., 1999*). Moreover, RNA interference on genes that regulate cell-cell adhesion and gap junction can significantly increase the positional variability between individuals (*Li et al., 2018*). The regulated or perturbed morphological phenotypes above can be simulated with our computational approaches.

Our framework can also be applied to study the characteristics of cell internalization discovered in other organismal systems (*Pinheiro and Heisenberg, 2020*; *Guo et al., 2021*). For example, in the simulation with experimental parameters, we identify 35 pairs of neighboring cells that internalize one after another at a specific stage, resembling the pattern of mesendoderm internalization during zebrafish gastrulation (*Table S10*) (*Pinheiro and Heisenberg, 2020*). To distinguish if the second internalizing cell is dragged inward by the first one, we remove the attractive force between them and find that all the second cells can still internalize, suggesting that at least in the situation of *C. elegans* embryo modeling, the consecutive internalization of neighboring cells is independent on the attraction between them.

Gastrulation is a fascinating biological phenomenon and little is known about the design principle on its regulatory programs, especially in mechanical and energetic considerations. About the *C. elegans* initial gastrulation, the activation timing *in vivo* corresponds to the predicted emergence of bistability while the regulated ingressing cells E2 has an elongated cell cycle and thus the largest cell size to facilitate their internalization. The feature match between *in vivo* and *in silico* indicates that the underlying genetic programs on gastrulation are optimized following the guidance of basic mechanics and energetics, to ensure developmental accuracy and minimize genetic instructions. In the future, the model on gastrulation could be improved by considering more biochemical cues (e.g., maternal-zygotic transition) and biophysical regulations (e.g., asymmetric adhesion and apical contraction) to fit the experimental conditions and study this morphogenetic process as well as other factors' functions more accurately (*Wong et al., 2016*; *Marston et al., 2016*). To sum up, this work not only provides a novel interpretation of worm gastrulation but also serves as an example of how to rationally interpret the genetic programs that have been optimized over the long history of evolution, from the mechanical and energetic perspectives.

**MATERIALS AND METHODS**

*Mechanical Model*

The motion of cells is modeled within an ellipsoidal shell. The lengths of three semi-axes of the ellipsoid are $L_x$ = 27 μm (anterior-posterior), $L_y$ = 18 μm (left-right), and $L_z$ = 18 μm (dorsal-ventral) according to a recent measurement on the size of 17 *C. elegans* wild-type embryos (*Cao et al., 2020*). A lot of physical models have been proposed to reconstruct the *C. elegans* embryonic morphology *in silico*, including the multi-particle model (*Kajita et al., 2002; Kajita et al., 2003*), coarse-grained model (*Fickentscher et al., 2013; Fickentscher et al., 2016; Yamamoto and Kimura, 2017; Tian et al., 2020; Guan et al., 2020*), and phase field model (*Jiang et al., 2019; Kuang et al., 2020*). Among them, the coarse-grained ones have shown a great advantage in scaling up the cell number and capturing the fundamental characteristics of morphogenetic dynamics.

Here, we follow the previous work which improved the coarse-grained model by optimizing the system parameters using *in vivo* embryo morphology data (*Guan et al., 2020; Cao et al., 2020*). The pressure exerted on a cell from the shell is given by

$$\boldsymbol{F}_{shell \to i} = k_{shell} \boldsymbol{e}_{shell \to i} \cdot \begin{cases} R_i - d_i, & 0 < d_i \leq R_i \\ 0, & d_i > R_i \end{cases} \quad (1)$$

where $k_{shell}$ is the elastic coefficient representing the intensity of cell-shell repulsion; $\boldsymbol{e}_{shell \to i}$ is the direction vector orienting from the shell (contact point) to cell $i$; $d_i$ is the minimal distance between the shell and cell $i$ (*Eberly, 2019*); $R_i$ is the radius of cell $i$. Apart, the interaction between cells combines both repulsion (volume effect associated with cell stiffness when $r_{ij}$ is small) and attraction (membrane adhesion when $r_{ij}$ is large). For simplicity, the interaction is described as linear functions controlled by the mean radius of cell $i$ and cell $j$,

$$\boldsymbol{F}_{j \to i} = \frac{F_{cell}}{2\alpha \overline{R}_{ij}} \boldsymbol{e}_{j \to i} \cdot \begin{cases} 2\alpha \overline{R}_{ij} - r_{ij}, & 0 < r_{ij} \leq (1+\alpha)\overline{R}_{ij} \\ r_{ij} - 2\overline{R}_{ij}, & (1+\alpha)\overline{R}_{ij} < r_{ij} \leq 2\overline{R}_{ij} \\ 0, & r_{ij} > 2\overline{R}_{ij} \end{cases} \quad (2)$$

where $F_{cell} = k_{shell} K$ represents intercellular elastic capacity with $k_{shell}$ as unit; $\boldsymbol{e}_{j \to i}$ is the direction vector orienting from cell $j$ to cell $i$; $\overline{R}_{ij}$ is the mean radius of cell $i$ and cell $j$; $\alpha$ determines the range of repulsive force. $K = 0.01$ μm and $\alpha = 0.9$ were found to be optimal for the *C. elegans* embryo during 4- to ~50-cell stages. It's worth pointing out that, only the cells identified as Voronoi neighbors are taken into account.

Since the environment inside the shell is assumed to be highly viscous, the motion of a cell can be described by an overdamped Langevin equation (*Fickentscher et al., 2013*),

$$\zeta \frac{d\boldsymbol{r}_i}{dt} = \boldsymbol{F}_{shell \to i} + \sum_j \boldsymbol{F}_{j \to i} + \boldsymbol{f}_i \quad (3)$$

Here, $\zeta$ is the viscosity coefficient; $\boldsymbol{r}_i$ is the position vector of cell $i$; $\boldsymbol{f}_i$ is the random force with a mean $\langle \boldsymbol{f}_i \rangle = 0$ and a variance $\langle \boldsymbol{f}_i(t)\boldsymbol{f}_j(t') \rangle = M\delta_{ij}\delta(t-t')$, where $M$ is the noise amplitude, $\delta_{ij}$ is the Kronecker delta function, and $\delta(t-t')$ is the Dirac delta function. The discrete form of *Equation 3* reads

$$\boldsymbol{r}_i(t+\Delta t) = \boldsymbol{r}_i(t) + \Delta t[\boldsymbol{F}_{shell \to i}(t) + \sum_j \boldsymbol{F}_{j \to i}(t)] + \sqrt{\Delta t}\eta\boldsymbol{\xi} \quad (4)$$

where $\eta = \sqrt{\frac{M}{\zeta}}$ is the equivalent noise amplitude and $\boldsymbol{\xi}$ is a three-dimensional random variable obeying the standard normal distribution.

When it comes to the equilibrium state searching and active migratory regulation, we add a driving force on the specific cell, which is directed toward the inner side of the shell and is expressed as

$$\boldsymbol{F}_{driving} = \kappa F_{cell}\left(2 - \frac{d_i}{R_i}\right)\boldsymbol{e}_{shell \to i}, \quad 0 < d_i \leq 2R_i \quad (5)$$

where $\kappa$ represents the intensity of the force. In the case of equilibrium state searching (*Figure 5A-B*), the force is exerted on Ea or Ep cell at each developmental stage before its division, i.e., 14-, 15-, 23-, 24-, 26-, 28-, and 44-cell stages. Then the force is permanently removed when the distance between the shell and cell reaches 1.9 times the cell radius. We find the minimal force intensity that can lead Ea or Ep cell to internalize by scanning $\kappa$ from 0.1 to 5 with an interval of 0.1, and choose Ep for driven internalization eventually for that it's easier to internalize than Ea. In the case of active migratory regulation (*Figure 8A-D*), the force is exerted on both Ea and Ep cells since 26-cell stage to simulate the active driving force during gastrulation, which is caused by asymmetric adhesion and apical contraction (*Marston et al., 2016*).

The simulation is performed from 4- to 330-cell stages in accordance with the highly conserved cell division sequence (27 groups of division events in total) obtained from *in vivo* experiments (*Table S2*) (*Guan et al., 2019*). The configuration of the first 4 cells (i.e., ABa, ABp, EMS, and P2) is initialized as the well-known rhombic pattern (*Kajita et al., 2002*; *Kajita et al., 2003*). All the cells are assumed to undergo symmetric divisions except the P lineages (i.e., P0, P1, P2, and P3), whose volume segregation ratios are assigned based on experimental measurement. The volume segregation ratios of P0, P1, P2, and P3 are 1.324, 1.557, 2.045, and 2.364, respectively, while the last P cell, P4, is set to divide equally to generate the primordial germ cells Z2 and Z3, which keep arrested until larva stage. The cell division event is approximated by replacing each dividing cell with two daughter cells positioned along the prescribed division orientation, which have an initial distance of 1.16 times of the radius of the mother cell ($D_{\text{sister}} / R_{\text{mother}} = 1.16$; *Figure S4*). All the volume and distance ratios are obtained from the experimental data of 17 segmented *C. elegans* wild-type embryos collected in *Cao et al., 2020*. Each group of cell division events are activated when the whole system's motion perturbed by the previous ones reaches equilibrium, in other words, its mean kinetic energy is close to the fluctuation energy associated with the intrinsic noise:

$$\sum_{j=1}^{N_{step}} \sum_{i=1}^{N} \left| \frac{r_i(t_j+\Delta t) - r_i(t_j)}{\Delta t} \right|^2 / 2 / N / N_{step} < 10^{-8} + \frac{3\eta^2}{2\Delta t} \tag{5}$$

where $N$ is the total cell number; $N_{step} = 600000$ is the inspected time step number; $\Delta t$ is the time step and chosen to be 0.1; $\frac{3\eta^2}{2\Delta t}$ represents the fluctuation energy of stochastic motion and the small amount $10^{-8}$ demands the system to approach equilibrium with very little additional energy.

### *Division Orientation*

The orientation of cell division is autonomous and set by the following rules:

1. As a default, a cell lying at the position $P(x, y, z)$ divides along the direction vector $\boldsymbol{e}_{division} = (y\Delta_y + z\Delta_z, -y\Delta_x, -z\Delta_x)$, where $\boldsymbol{e}_{outward} = (\Delta_x, \Delta_y, \Delta_z)$ is the vector orienting from the position $P$ to the point of the shell that is the closest to $P$. Geometrically, $\boldsymbol{e}_{division}$ is perpendicular to $\boldsymbol{e}_{outward}$ and located in a plane overlapping the $x$ axis. This quantitative rule can resemble the actual cell division behaviors in which most division orientations are near the anterior-posterior axis and approximately tangent to the shell.

    Several exceptions are listed below to prevent the potential problems in cell identification and structural bifurcation during simulation:

2. When a cell lies nearby the $x$ axis (with a distance to the $x$ axis less than 0.2 times the cell radius) but off the anterior and posterior terminals (with a distance to the shell longer than 1.2 times the cell radius), it divides along the direction vector $\boldsymbol{e}_x$.

3. When a cell lies nearby the $x$ axis and around the anterior or posterior terminal (with a distance to the shell no longer than 1.2 times the cell radius), it divides along the direction vector $0.100\boldsymbol{e}_x - 0.995\boldsymbol{e}_z$.

4. Since the cells ABa and ABp in the wild-type embryos divide approximately along the left-right axis ($y$) (*Sugioka and Bowerman, 2018*), violating the default above, extra clockwise rotations of 45° are performed on their division orientations and the rotation axes are set as $-\boldsymbol{e}_x$ and $\boldsymbol{e}_z$ for ABa and ABp respectively.

*Cell Identification*

The cell identity/name is determined by its mother's division orientation according to the nomenclature in *C. elegans* embryogenesis research (*Sulston et al., 1983*). Generally, for the symmetric division of a cell *X*, the two daughters are named as *X*a (closer to $-e_x$) and *X*p (closer to $e_x$). Note that there are three exceptions: (1) for the germline stem cell divisions, P0, P1, P2, P3, and P4's daughters are named as P1, P2, P3, P4, and Z3 (closer to $e_x$) and AB, EMS, C, D, and Z2 (closer to $-e_x$), respectively. (2) for the ABa and ABp cells, the daughters are named as *X*l (closer to $-e_y$) and *X*r (closer to $e_y$); (3) for the EMS cell, the daughters are named as E (closer to $e_x$) and MS (closer to $-e_x$).

**ACKNOWLEDGMENTS**

We thank Siyu Chen, Prof. Xiaojing Yang, and all the members of Tang Lab for helpful discussions and comments. This work was supported by the National Natural Science Foundation of China (Grant Nos. 12090053, 32088101). Part of the computation was carried out on the High-Performance Computing Platform of the Center for Life Sciences at Peking University.

**COMPETING INTERESTS**

The authors declare that no competing interests exist.

**AUTHOR ORCIDS**

Jiao Miao, https://orcid.org/0000-0003-3097-1281

Guoye Guan, https://orcid.org/0000-0003-4479-4722

Chao Tang, https://orcid.org/0000-0003-1474-3705

**AUTHOR CONTRIBUTIONS**

Jiao Miao, Guoye Guan, Conceptualization, Data curation, Formal analysis, Investigation, Methodology, Software, Validation, Visualization, Writing - original draft, Writing - review and editing; Chao Tang, Conceptualization, Formal analysis, Funding acquisition, Methodology, Project administration, Resources, Supervision, Validation, Writing - review and editing.